\documentclass[12pt,preprint]{aastex} 

\slugcomment{Submitted to The Astronomical Journal}
 
\shortauthors{Clampin {\it et al.}} 
\shorttitle{{\itHST}/ACS Imaging of HD 141569} 

\begin{document} 
\title{{\it HST}/ACS Coronagraphic Imaging of the \\Circumstellar Disk around HD~141569A} 

\author{
M.~Clampin\altaffilmark{1}, 
J.~E.~Krist\altaffilmark{1},
D.~R.~Ardila\altaffilmark{2}, 
D.~A.~Golimowski\altaffilmark{2},
G.~F.~Hartig\altaffilmark{1}, 
H.~C.~Ford\altaffilmark{2}, 
G.~D.~Illingworth\altaffilmark{3},
F.~Bartko\altaffilmark{4}, 
N.~Benitez\altaffilmark{2}, 
J.~P.~Blakeslee\altaffilmark{2}, 
R.~J.~Bouwens\altaffilmark{3}, 
T.~J.~Broadhurst\altaffilmark{5}, 
R.~A.~Brown\altaffilmark{1}, 
C.~J.~Burrows\altaffilmark{1},
E.~S.~Cheng\altaffilmark{6}, 
N.~J.~G.~Cross\altaffilmark{2},
P.~D.~Feldman\altaffilmark{2}, 
M.~Franx\altaffilmark{7},
C.~Gronwall\altaffilmark{2},
L.~Infante\altaffilmark{8},
R.~A.~Kimble\altaffilmark{7}, 
M.~P.~Lesser\altaffilmark{9}, 
A.~R.~Martel\altaffilmark{2}, 
F.~Menanteau\altaffilmark{2},
G.~R.~Meurer\altaffilmark{2}, 
G.~K.~Miley\altaffilmark{7}, 
M.~Postman\altaffilmark{1},
P.~Rosati\altaffilmark{10}, 
M.~Sirianni\altaffilmark{2},  
W.~B.~Sparks\altaffilmark{1},
H.~D.~Tran\altaffilmark{2}, 
Z.~I.~Tsvetanov\altaffilmark{2},  
R.~L.~White\altaffilmark{1},
and W.~Zheng\altaffilmark{2}
} 

\altaffiltext{1}{STScI, 3700 San Martin Drive, Baltimore, MD 21218.}
\altaffiltext{2}{Department of Physics and Astronomy, Johns Hopkins University,\\
  Baltimore, MD 21218.}
\altaffiltext{3}{Lick Observatory, University of California, Santa Cruz, CA 95064.}
\altaffiltext{4}{Bartko Sci. \& Tech., P.O. Box 670, Mead, CO 80542-0670.}
\altaffiltext{5}{The Racah Institute of Physics, Hebrew University, Jerusalem \\
  91904, Israel.}
\altaffiltext{6}{NASA-GSFC, Greenbelt, MD 20771.}
\altaffiltext{7}{Leiden Observatory, P.O. Box 9513, 2300 Leiden, The 
  Netherlands.}
\altaffiltext{8} {Pontificia Universidad Catolica de Chile, Santiago, Chile.}
\altaffiltext{9}{Steward Observatory, University of Arizona, Tucson, AZ 85721.}
\altaffiltext{10}{European Southern Observatory, Karl-Schwarzschild-Str. 2, \\
  D-85748 Garching, Germany.}

\begin{abstract} Multicolor coronagraphic images of the circumstellar disk
around HD~141569A have been obtained with the {\it Hubble Space Telescope
(HST)} Advanced Camera for Surveys.  $B$, $V$, and $I$ images show that the
disk's previously-described multiple-ring structure is actually a continuous
distribution of dust with a tightly-wound spiral structure.  Two, more open
spiral arms extend from the disk, one of which appears to reach the nearby
binary star HD~141569BC.  Diffuse dust is seen up to 1200~AU from HD~141569A.
Although planets may exist in the inner region of the disk, tidal interaction
with HD~141569BC seems more likely to be the cause of these phenomena. The
disk appears redder than the star ($B$--$V = 0.21$ and $V$--$I = 0.25$), 
and its color is spatially uniform.  A
scattering asymmetry factor of $g = 0.25$--0.35 is derived.  The azimuthal
density distribution is asymmetric, varying by a factor of $\sim 3$ at some
radii.  \end{abstract}

\keywords{stars: circumstellar matter --- stars: individual (HD~141569A)
--- stars: pre-main sequence}

\section{Introduction}

HD~141569A (spectral type B9.5Ve; Jaschek \& Jaschek 1992) is one of a few stars 
from which excess infrared radiation has been detected by IRAS and for which an 
associated circumstellar debris disk has been imaged in reflected light.  Although
HD~141569A was initially classified as an A-type shell star with infrared excess
(Jaschek, Jaschek, \& Egret 1986), it has properties common to both Herbig Ae/Be
stars and young main-sequence stars (Fisher et al.\ 2000).

Images of HD~141569A's circumstellar disk show complex structure (Weinberger et
al.\ 1999, hereafter W99; Augereau et al.\ 1999, hereafter A99) that might be
attributed to gravitational perturbation by one or more planets (W99; Mouillet
et al.\ 2001, hereafter M01).  Other studies conclude that the 5~Myr-old disk
may be too young to have formed Jovian planets (Weinberger et al.\ 2000) and
have proposed alternative sculpting mechanisms, such as dust migration
(Takeuchi \& Artymowicz 2001).  The disk has a fractional infrared excess
luminosity of $L_{\rm disk} / L_* = 8.4 \times 10^{-3}$ -- three times that of
$\beta$ Pic's disk (Sylvester et al.\ 1996) -- and has been imaged to a radius
of $\sim 500$~AU.  At a distance of 99 pc (ESA 1997), the disk has 
has an angular radius of $5''$ and is therefore an ideal target for imaging
with {\it HST}.

{\it HST's} Advanced Camera for Surveys (ACS) High Resolution Channel (HRC)
features a coronagraph that suppresses the wings of the point-spread function 
(PSF) by factors of 5 to 10.  When combined with reference PSF subtraction, 
the contrast between faint circumstellar material and the scattered light from 
a bright star can be improved by factors of 500--1000 (Krist et al.\ 2002) at 
a radial distance of 2$\arcsec$.  The HRC coronagraph images of HD~141569A presented
here show the complex structure of the disk with unprecedented resolution and provide new information about the disk's color.

\section{Observations and Data Reduction}

HD~141569A was observed with the HRC coronagraph on UT 2002 July 21 as part of the ACS Early Release Observation (ERO) program.  The angular size 
of an HRC pixel is 0\farcs028$\times$ 0\farcs025, and the FWHM of the unocculted
coronagraphic field PSF is 0\farcs050 in the $V$ band.  The star was positioned
behind the 1\farcs8-diameter occulting spot.  A sequence of images was recorded
over two orbits, after which {\it HST} was rolled by 28$^{\circ}$ and the 
sequence was repeated.  This roll offset permits the distinction of PSF features,
which are stationary with roll, from real disk structure, whose orientation on
the detector changes with roll.  Each sequence comprised the following filters 
and exposures: F435W ($B$), $1 \times 150$~s, $3 \times 760$~s; F606W (broad $V$),
$1 \times 100$~s, $3 \times 685$~s; F814W ($I$), $1 \times 100$~s, $2 \times 
1200$~s.  The long exposures were not saturated, so the short ones were not used
in our analysis.  

The images were reduced using the standard techniques of bias and dark subtraction
and division by a flat field.  Because the coronagraphic spots shifted during 
launch, the pre-launch coronagraphic flat fields could not be applied directly
to our images.  Moreover, on-orbit flat fields were not available.  
Consequently, we constructed suitable flat fields by shifting the coronagraphic
illumination pattern (including the occulting spot shadows) within the pre-launch
flat fields.  The flattened images were then combined using a standard cosmic-ray
rejection algorithm.

The calibrated images of HD~141569A reveal the presence of the disk, but the 
coronagraphic PSF still dominates the images (Figure~1).  The coronagraph 
effectively suppresses the diffracted light below the level of scattered light
caused by mid-frequency surface errors in {\it HST's} mirrors.  The occulting 
spot itself diffracts light into a concentrated area around the perimeter of
the imaged spot.  Because the coronagraphic PSF is stable, most of this residual
light can be subtracted using an image of a suitable reference star,
improving the contrast.

Techniques for subtracting HRC coronagraphic PSFs are described by Krist et al.\ 
(2002) and summarized in the ACS Instrument Handbook (Pavlovsky et al.\ 2002).  
Because the positions of the occulting spots in the image plane are not static, 
optimal PSF subtraction can be achieved only when images of the target and 
spectrally-similar reference stars are recorded contemporaneously.  Previous studies of HD~141569A's disk 
with NICMOS (W99; A99) and STIS (M01) employed PSF subtraction, usually with a
single reference star.  W99 used three reference stars observed at various times and
selected the one that provided the best subtraction.  Because the ERO program produced 
the first on-orbit images taken with the HRC coronagraph, we are limited to using one set 
of reference-star images for PSF subtraction. 

We observed the reference star HD~129433 
immediately after HD~141569A.  Both stars
have the same spectral type, but HD~129433 is less reddened ($B$--$V = -0.01$, 
versus 0.09).  The images of HD~129433 were normalized to, aligned with, and 
subtracted from the corresponding images of HD~141569A (Figures~1 and 2).  
Normalization and alignment of the HD~129433 images were achieved by simultaneously 
and iteratively scaling and shifting using cubic-convolution interpolation.  
Convergence was achieved when the subtraction residuals were visually minimized.  
Alignment of the images is accurate to within $\pm 0.05$~pixel ($\pm$~0\farcs0013). 
The normalizations of the corresponding images are accurate to within 2\%.  This 
normalization error corresponds to photometric errors of 0.02~mag~arcsec$^{-2}$ 
and 0.1~mag~arcsec$^{-2}$ in the brightest and faintest regions of the disk, 
respectively.With only one available reference star, we cannot quantify the 
subtraction errors caused by mismatches of the stars' colors or time-dependent
PSF variations.

The largest residuals in each PSF-subtracted image appear at the perimeter of the 
occulting spot, where the PSF is most sensitive to star-to-spot registration.
These residuals force the exclusion of data within $\sim 1\farcs2$ (120~AU) of the 
star.  Radial streaks caused by mismatches in registration, color, and focus also
appear in the subtracted images.  The F814W images are the most affected by 
residuals, probably because of large star-to-spot misalignments.  After correcting 
for geometric distortion and rotating to a common orientation, the images for each 
filter were combined so that the final images reflected the smallest subtraction 
residuals from each of the constituent images (Figure~3).

To derive the optical depth of HD~141569A's disk, the images must be normalized by the stellar flux.  Unfortunately, it was not possible to obtain unsaturated images of HD~141569A and HD~129433 through HRC broadband filters.  Consequently, we computed the flux from each star through each filter using the STSDAS synthetic photometry package {\small SYNPHOT}, which simulates most {\it HST} observing configurations.  To approximate HD~129433's spectrum, we normalized Vega's spectrum to $V=5.73$ (Hoffleit \& Jaschek 1982) with $A_V=0$ and assumed invariability.  We also used Vega's spectrum for HD~141569A with $A_V=0.34$ (Oudmaijer et al.\ 2001).  Because HD~141569A is a known variable, we scaled its spectrum so that the ratio of the synthetic F606W fluxes of HD~141569A and HD~129433, $f_{\rm F606W}$, was equal to the average brightness ratio of the aligned F606W coronagraphic PSFs of the two stars.  The resulting magnitude for HD~141569A, $V=7.18$, matches well the published measurement of de~Winter et al.\ (2001).  We then compared the ratios of the stars' flux-ratios through different filters (e.g., $f_{\rm F435W}/f_{\rm F814W}$ and $f_{\rm F606W}/f_{\rm F814W}$) computed from the coronagraphic PSFs and from the synthetic spectra.  The actual and synthetic flux ratios differed by less than 1\%.  This agreement indicates that the broadband colors of the synthetic spectra are well matched to the stars' colors, and that the spectra can be used to produce accurate synthetic fluxes through the HRC filters.  The PSF-subtracted images of the disk were normalized to the synthetic fluxes of HD~141569A, assuming a 52.5\% throughput reduction caused by the coronagraph's Lyot stop.

\section{Results}

Our PSF-subtracted images yield a clearer and more detailed view of the disk
than seen in previous {\it HST} images obtained with NICMOS (W99; A99) and STIS
(M01).  The disk has four distinct annular zones: an inner clearing within
$\sim 175$~AU of the star, a bright ``ring'' with a sharp inner edge from $\sim
175$~AU to $\sim 215$~AU, a faint zone from $\sim 215$~AU to $\sim 300$~AU, and
a broad ``ring'' from $\sim 300$~AU to $\sim 400$~AU.  (All distances are measured 
along the disk's southern semimajor axis.) As shown in Section~4, the
inner and outer rings appear to be thin and tightly wound spirals of dust.
Faint, broad arcs superposed upon fainter and more diffuse dust extend from the
northeast and southwest regions of the disk (Figure~4).  The diffuse dust in
the northeast extends $\sim 1200$~AU from HD~141569A.  The arcs resemble open
spiral arms, and the southwest one extends toward the binary star HD~141569BC
located $\sim$~8\farcs5 to the northwest of HD~141569A.

Figure~4 also shows two faint stars, labeled 1 and 2, located 6\farcs3 and 
8\farcs0 from HD~141569A at position angles 12.5$^{\circ}$ and 210.0$^{\circ}$, 
respectively.  Their $V$ magnitudes are $25.5 \pm 0.1$ and $24.0 \pm 0.1$, 
respectively.  Their colors are consistent with slightly reddened K stars or 
extincted, earlier-type stars, and we conclude that they are background stars.

Figures 5 and 6 show the $V$-band surface brightness of the disk, derived from
the F606W images after azimuthal filtering (described below).  The surface
brightness ranges from $16.5 \pm 0.02$~mag~arcsec$^{-2}$ in the inner ring to
$21.5 \pm 0.1$~mag~arcsec$^{-2}$ at the furthest extent of detection.  Large
azimuthal variations in the surface brightness of the disk are evident; the
brightness of the outer ring varies by a factor of 2.5.  The F814W/F606W and
F814W/F435W ratios of the average surface brightness of the disk are $10 \pm
0.7$\% and $25 \pm 2$\% larger, respectively, than the corresponding flux
ratios of HD~141569A.  These ratios correspond to $B$--$V = 0.21$ and $V$--$I = 0.25$, 
which indicate that
the disk is significantly redder than the star. There is no evidence of
color variation across the disk above the levels set by the local
PSF-subtraction residuals.

The north--south orientation of the semimajor axis of the projected disk and the 
likely condition of forward scattering by dust (W99) imply that the brighter, eastern side
of the disk is nearer to us.  To investigate the scattering properties further, we
created a composite image of the disk by summing the F435W and F606W images after 
normalizing them to a common average surface brightness.  (Because the F814W image
has relatively large PSF-subtraction residuals, it was not included in the 
composite image.)  Assuming that the disk is optically thin and flat, and that it 
has an inclination angle of 55$^{\circ}$ from pole-on (M01), we deprojected the 
composite image to simulate a face-on view of the disk (Figure~7a).  The radial 
subtraction residuals seen in the constituent images (Figure~2) were diminished by 
azimuthally filtering the combined image (Figure~7b).  Each pixel was replaced with 
the median value of a seven-pixel arc centered on the geometic center used in the 
deprojection procedure.  This technique exploits the disk's slow azimuthal
variations while preserving its more rapid radial ones.  Each pixel in the
azimuthally filtered image was then multiplied by the square of its distance
from HD~141569A to compensate for the radially diminishing stellar illumination
in the optically-thin disk (Figure~7c).  The enhanced brightness of the near
side of the disk and its symmetry along the line of sight as shown in this image
indicates significant forward scattering of starlight by the dust.

W99 derived a scattering asymmetry parameter of $g=0.11$ (Henyey \& Greenstein
1941) for the disk based on a measured east/west brightness ratio of $1.5 \pm
0.2$ and an assumption of azimuthally uniform disk density distribution.  However, our
images show that the disk is not azimuthally uniform.  Consequently, we
visually fitted a Henyey--Greenstein phase function to our entire deprojected F435W +
F606W image,  accounting for the inclination.  We adjusted $g$ until the region
of symmetrical brightness enhancement along the disk's near side was made
consistent with the overall azimuthal brightness trend (Figure~7d).  We
obtained satisfactory results for $g=0.25-0.35$, which is consistent
with the results of M01.  Because the color of the disk appears uniform, this
range of $g$ is valid for all visual wavelengths.  Depending on the infrared 
properties of the dust, $g$ may decrease for wavelengths longer than $1.0~\mu$m.
For example, isotropic scattering ($g=0$) would explain the brighter western side 
of the disk in $1.6~\mu$m images obtained with NICMOS (A99).   
 
Figure~7d maps the product of the dust's albedo ($\omega$) and its optical
depth perpendicular to the plane of the disk ($\tau_\perp$). This is proportional
 to $\kappa_s\Sigma$, where $\kappa_s$ is the scattering opacity
and $\Sigma$ is the surface mass density of the dust. The map of
$\omega\tau_\perp$ enhances the features within the disk, especially the
tightly-wound spiral structure of the inner and outer rings.  These tight
spirals and the broad, open spiral arms extending beyond the outer ring are
traced in Figure~8.  The outer ring unwinds counterclockwise, circumscribing
about $450^{\circ}$ of arc.  The spiral overlaps itself in the northeast with a
separation of $\sim 80$~AU.  It also splits into the open spiral arms in the
northeast and southwest.  The inner ring also unwinds counterclockwise and
overlaps itself from northeast to southeast.  This spiral appears to merge with
the outer spiral in the southeast, thereby filling the previously described gap
between the inner and outer rings about 250~AU from the star.  The interior of
the inner ring appears relatively devoid of dust.  

Figure~9 is a color-coded version of Figure~7d with an expanded field of view
and a calibrated brightness scale.  Assuming $\kappa_s$ is uniform throughout the disk
(consistent with the observed color uniformity), $\Sigma$ is $\sim 3$ times greater in
the southwest region of the outer ring than in the northeast.  The
peak of the apparent optical depth in the outer ring is $\omega\tau_\perp\approx 0.006$,
assuming $g = 0.25$.  W99 derived a mean $\omega\tau_\perp\approx 0.0036$ 
for the entire outer ring without correcting for forward scattering.    

Because the HRC's occulting spots lie in the spherically aberrated beam from
{\it HST}, some aberrated starlight passes by the spots and is corrected by the
HRC optics. This corrected light forms a diminished image of the star on the detector 
within the imaged shadow of the spot (Figure~1), which allows direct measurement 
of the star's position. (This technique was verified during post-launch calibration of ACS.)
Our images suggest that HD~141569A is not located at the center of the circumstellar disk. However, 
the disk's spiral structure precludes an accurate determination of its center. The offset 
appears to be of order $\sim$30 AU toward the west side of the disk, but the uncertainty in the 
measurement could exceed $\sim$10 AU.  An offset is also reported in STIS images of the disk 
(G.~Schneider 2002, private communication).  M01 did not report a star-to-disk offset in
their STIS images, but they did note that the centers of the inner and outer
rings are offset by 0.2~AU.  We do not see an offset between the rings we show in Figure~8 although, once
again, the spiral structure complicates such a measurement.  

\section{Discussion}

Spiral structure has been observed in the circumstellar disk around HD~100546
(Grady et al.\ 2001), but that disk does not possess the open and extended
spiral arms seen in HD~141569A's disk.  The presence of these arms, the
apparent association of the southwest arm with HD~141569BC, the large azimuthal
density variations of the disk, and the $\sim 25$~AU offset of HD~141569A
relative to the center of the disk are characteristics consistent with the
tidal effects of encounters between circumstellar disks and passing stars
(Pfalzner, Henning, \& Kley 2000; Larwood \& Kalas 2001).  If HD~141569A and
HD~141569BC are bound, then a few hundred encounters between the disk and
HD~141569BC could have occurred over the 5~Myr age of the system.  Under such
circumstances, the disk would be tidally truncated at a distance of $\sim
0.2$--0.4 times the semimajor axis of the orbit and its outer regions would
develop spiral structure (Artymowicz \& Lubow 1994).  The eccentricity of the
HD~141569A/BC system would also induce eccentricity in the disk that could
account for the displacement of the disk with respect to HD~141569A.  Detailed
dynamical modeling of the system is required to explore fully the possible
interaction of the disk with HD~141569BC.

The redness of the disk relative to HD~141569A indicates that the grain
properties are different from those of the neutrally-colored but older disk
around $\beta$~Pictoris.  However, the color of HD~141569A's disk is similar to
HR~4796A's disk (Schneider \& Silverstone 2002) and indicates that the scattering
efficiency of the dust increases with wavelength.  While our images cannot
strongly constrain the distribution and size of the dust grains, they do 
permit quantitative speculation.  For example, a reasonable match to the observed
colors is obtained if we assume that the grains are composed of 
``astronomical silicate'' (Laor \& Draine 1993; Draine \& Lee 1984), and that the 
distribution of grain sizes obeys a power law with an index of 3.5 and radii extrema of
$0.4~\mu$m and $10~\mu$m.  These assumptions imply a $V$-band albedo of $\omega\approx 0.6$.  
Other combinations of the grain composition, size distribution, and albedo may, however, 
produce equally good matches to the observed disk colors.  Nevertheless, the size distribution quoted above
is consistent with that derived by A99 from the disk's 8--$100~\mu$m spectral energy
distribution and an assumed composition of silicate grains, organic refractories,
and water ice.

The spatial uniformity of the disk's color does not necessarily imply that the
distribution of grain sizes is constant throughout the disk.  The colors are
not sensitive to the distribution of grains with sizes $\gtrsim 5~\mu$m.
Therefore, color uniformity alone is not inconsistent with the scenario of dust
segregation in the presence of stellar radiation pressure (Takeuchi \&
Artymowicz 2001).  However, the azimuthally varying density distribution of the
grains (Figure~9) and the protracted time scale of such segregation relative to
possible periodic encounters with HD~141569BC make dust segregation an unlikely
cause of the observed disk structure. 
 
The complex structure of the disk prompts us to reconsider the potential
influence of a very-low-mass companion on the disk's morphology.  W99 have
suggested that the depleted region of the disk from $\sim 215$--300~AU could be
produced from tidal clearing by a Jupiter-sized planet in the gap itself.
Takeuchi \& Artymowicz (2001) and Klahr \& Lin (2002) have argued that the arcs
seen in the disks around HR~4796A and HD~141569A may not necessarily be
indicators of embedded planets.  Current theories suggest that planet formation
at such large distances from HD~141569A is unlikely within 5~Myr (Boss 1998).
Moreover, Takeuchi \& Artymowicz (2001) argue that the creation of a circular
depleted region by an outwardly migrating planet is unlikely.  Alternatively,
HD~141569BC may have induced the scattering of a planet from a small orbit to
the gap.  Thommes, Duncan, \& Levison (2002) have also shown that planets
within 100~AU may scatter objects out to 300~AU.  Whether the orbit of a
scattered planet within HD~141569A's disk could circularize within 5~Myr is
unclear.  On the other hand, the cause of the gap need not be in the gap
itself; a planet in a close orbit might clear the gap via excited density waves
within its outer Lindblad resonance (Goldreich \& Tremaine 1978).  Indeed,
Brittain \& Rettig (2002) have presented evidence for H$_3^+$ within 17~AU of
HD~141569A, which may be attributed to the extended envelope of a protoplanet.
All these possibilities remain to be investigated.  Our images do not exclude a
planet as the mechanism responsible for some of the disk's structure.

The causes of the complicated structures observed in the dust disks around 
HD~141569A and other young stars clearly require further investigation and modeling.
As our images demonstrate, the ACS coronagraph can produce images of these disks 
with unprecedented resolution and contrast.  Such high-quality images permit the 
type of studies necessary to address quantitatively the many outstanding questions 
regarding the composition, dynamics, and evolution of protoplanetary disks.

\acknowledgements

We are grateful to P.~Artymowicz, S.~Lubow, J.~Pringle and D. Lin for 
discussions about
dynamical modelling, to G.~Schneider for details of prior {\it HST}/STIS
observations, and to H. Throop for help with the dust scattering models. 
We also thank T.~Allen, K.~Anderson, S.~Barkhouser, S.~Busching,
A.~Framarini, D.~Magee, and W.~J.~McCann for their invaluable contributions
to the ACS Investigation Definition Team (IDT) and Christine Klicka for assistance with the graphics.
This work was supported by NASA grant NAG5-7697 to the ACS IDT.

\pagebreak

\pagebreak

\begin{figure}
\epsscale{0.5}
\caption{
{\it (a)} HRC coronagraph image of HD~141569A through F606W without PSF subtraction
but corrected for geometric distortion.
{\it (b)} Similar image of reference star HD~129433.
{\it (c)} Image of HD~141569A's circumstellar disk after PSF subtraction. 
The images are displayed using a square-root stretch.}
\end{figure}

\begin{figure}
\epsscale{1.0}
\caption{
PSF-subtracted images of the disk obtained with three filters and two {\it HST} roll
orientations.  Images recorded at Roll~2 have been rotated to the same orientation 
as the images recorded at Roll~1.  The images are displayed using a square-root 
stretch.} 

\end{figure}

\begin{figure}
\epsscale{0.5}
\caption{
Images of the circumstellar disk after combining PSF-subtracted images from each 
roll orientation. The images are displayed using a square-root stretch.  Regions 
containing large subtraction residuals are masked.  The two black lobes in the 
F606W image mask residuals from the diagonal image artifact seen in Figure~1
and discussed by Krist et al.\ (2002).  The cross marks the measured position of 
the star.}
\end{figure}

\begin{figure}
\epsscale{0.5}
\caption{
PSF-subtracted, combined F435W+F606W image of the disk, stretched to
enhance the faint material extending from the front and back of
the disk.  Background stars 1 and 2 are labeled.}
\end{figure}

\begin{figure}
\epsscale{0.5}
\caption{
Color-coded map of the $V$-band surface brightness of the disk, obtained from the 
azimuthally filtered F606W image and photometrically calibrated from the 
synthetically derived flux of HD~141569A in this band.  
HD~141569BC lies in the upper left corner.  Background stars are
labeled 1 and 2. Note that the spatial scale is different from that of Figure~4.
}
\end{figure}

\begin{figure}
\epsscale{0.6}
\caption{
$V$-band surface brightness profiles along the major and minor axes of the disk, 
obtained from the azimuthally filtered F606W image and photometrically calibrated 
from the synthetically derived flux of HD~141569A in this band.  }
\end{figure}

\begin{figure}
\epsscale{0.5}
\caption{
Sequence of images illustrating the derivation of a map of the disk's vertical
optical depth.  Earth is toward the bottom and North to the left.  {\it (a)}
Deprojected F435W+F606W image of the disk (square-root stretch). 
{\it (b)} Deprojected image after azimuthal median filtering (square-root stretch).
{\it (c)} Azimuthally filtered image scaled to compensate for the inverse-square
drop in illumination by HD~141569A (linear stretch). 
{\it (d)} Image divided by a Henyey-Greenstein scattering asymmetry function with 
$g=0.25$ (linear stretch).  This image is proportional to the albedo $\times$ the 
optical depth perpendicular to the disk.}
\end{figure}

\begin{figure}
\epsscale{0.5}
\caption{{\it(a)} Surface density with traced spirals: inner (green),
outer (red), and open (blue). {\it (b)} Same image without traces.
 The disk orientation is the same as shown in Figure 7.}
\end{figure}

\begin{figure}
\epsscale{0.5}
\caption{
Color-coded map of the apparent optical depth (i.e., the product of the dust albedo, $\omega$,
and vertical optical depth, $\tau_\perp$) of the HD~141569A disk.  The image is normalized
to the peak value in the outer ring ($\omega\tau_{max}=0.006)$.  The disk orientation is the 
same as shown in Figure~7.  The cross marks the measured position 
of the star.  The corrections for the $r^{-2}$ illumination function 
and the scattering asymmetry make conspicuous the extended
spiral structure of the disk.}
\end{figure}

\end{document}